# Localization of protein aggregation in Escherichia coli is governed by diffusion and nucleoid macromolecular crowding effect

Anne-Sophie Coquel[1,3,4], Jean-Pascal Jacob[5], Mael Primet[5], Alice Demarez[1], Mariella Dimiccoli[5], Thomas Julou[6], Lionel Moisan[5], Ariel B. Lindner[1,2*], Hugues Berry[3,4*]

1. Institut National de la Santé et de la Recherche Médicale, Unité 1001, 75014 Paris, France
2. Faculty of Medicine, Paris Descartes University, 75014 Paris, France
3. EPI Beagle, INRIA Rhone-Alpes, 69603 Villeurbanne, France.
4. University of Lyon, LIRIS UMR5205 CNRS, 69621 Villeurbanne, France
5. University Paris Descartes, MAP5 - CNRS UMR 8145, 75270 Paris, France
6. Laboratoire de Physique Statistique de l'École Normale Supérieure, UMR 8550 CNRS, 24 rue Lhomond, 75005 Paris, France

*Equal contribution; corresponding authors: ariel.lindner@inserm.fr; hugues.berry@inria.fr

# ABSTRACT

Aggregates of misfolded proteins are a hallmark of many age-related diseases. Recently, they have been linked to aging of *Escherichia coli (E. coli)* where protein aggregates accumulate at the old pole region of the aging bacterium. Because of the potential of *E. coli* as a model organism, elucidating aging and protein aggregation in this bacterium may pave the way to significant advances in our global understanding of aging. A first obstacle along this path is to decipher the mechanisms by which protein aggregates are targeted to specific intercellular locations.

Here, using an integrated approach based on individual-based modeling, time-lapse fluorescence microscopy *and* automated image analysis, we show that the movement of aging-related protein aggregates in *E. coli* is purely diffusive (Brownian). Using single-particle tracking of protein aggregates in live *E. coli* cells, we estimated the average size and diffusion constant of the aggregates. Our results evidence that the aggregates passively diffuse within the cell, with diffusion constants that depend on their size in agreement with the Stokes-Einstein law. However, the aggregate displacements along the cell long axis are confined to a region that roughly corresponds to the nucleoid-free space in the cell pole, thus confirming the importance of increased macromolecular crowding in the nucleoids. We thus used 3d individual-based modeling to show that these three ingredients (diffusion, aggregation and diffusion hindrance in the nucleoids) are sufficient and necessary to reproduce the available experimental data on aggregate localization in the cells. Taken together, our results strongly support the hypothesis that the localization of aging-related protein aggregates in the poles of *E. coli* results from the coupling of passive diffusion-aggregation with spatially non-homogeneous macromolecular crowding. They further support the importance of "soft" intracellular structuring (based on macromolecular crowding) in diffusion-based protein localization in *E. coli*.

# AUTHOR SUMMARY

Localization of proteins to specific positions inside bacteria is crucial to several physiological processes including chromosome organization, chemotaxis or cell division. Since bacterial cells do not possess internal sub-compartments (e.g., cell organelles) nor vesicle-based sorting systems, protein localization in bacteria must rely on alternative mechanisms. In many instances, the nature of these mechanisms remains to be elucidated. In *Escherichia coli*, the localization of aggregates of misfolded proteins at the poles or the center of the cell has recently been linked to aging. However, the molecular mechanisms governing this localization of the protein aggregates remain controversial. To identify these mechanisms, we have devised an integrated strategy combining innovative experimental and modeling approaches. Our results show the importance of the increased macromolecular crowding in the nucleoids, the regions within the cell where the bacterial chromosome preferentially condensates. They indicate that a purely diffusive pattern of aggregates mobility combined with nucleoid occlusion underlies their accumulation in polar and mid-cell positions.

# INTRODUCTION

While aging is a fundamental characteristic of living systems, its underlying principles are still to be fully deciphered. Recent observations of ageing in unicellular models, in absence of genetic or environmental variability, have paved way to new quantitative experimental systems to address ageing's underlying molecular mechanisms [1,2]. Further, the notion of aging was extended beyond asymmetrically dividing unicellular organisms such as the budding yeast *Saccharomyces cerevisiae* or the bacterium *Caulobacter crescentus* -where a clear morphological difference and existence of a juvenile phase distinguishes between the aging mother cell and its daughter cells [3,4]- to symmetrically dividing bacteria. This pushed aging definition to demand *functional* asymmetry as minimal requirement for a system to age [5]. Specifically, *Escherichia coli* and *Bacillus subtilis* were shown to age as observed by loss of fitness at small generation scale (<10)   [6-8 (for B. *subtilis*),9-11] and increased probability of death at larger generation scale (up to 250 generations) [12]. Age in this system was defined as the number of consecutive divisions a cell has inherited the older cellular pole [7]; the sibling that inherits the older cell pole was shown to grow slower than the newer pole sibling.

From a cellular viewpoint, aging is arguably due to the accumulation of damage over time that degenerates cellular functions, ultimately affecting the survival of the organism [1,2]. In the case of *E. coli,* a significant portion of the age-related fitness loss is accounted for by the presence of protein aggregates that accumulate in the bacterial older poles [7,9,10]. Such accumulation is reminiscent of many known age-related protein folding diseases [1]. Preferential sequestration of damaged proteins is also observed in *S. cerevisiae* between the bud and the mother cell [14-16] and between specific intracellular compartments in yeast and mammalian cell [17,18]. Therefore spatial localization, as non-homogeneous distribution of damaged protein aggregates in the cytoplasm, has been postulated to be an optimized strategy allowing cell populations to maintain large growth rates in the face of the accumulation of damages that accompany metabolism during cell life [15,19,20]. These results suggest that spatial localization of damaged protein aggregates could present an ageing process conserved across different living kingdoms. Given the documented link between protein aggregation and ageing, the short life-span, ease of quantification of large number of individuals, molecular biology and genetics accessibility of *E. coli* may make this bacterium into a relevant model system to elucidate protein aggregation role in a ageing.

A first obstacle along this path is to understand the mechanisms by which cells can localize protein aggregates at specific locations within their intracellular space. Generally, thermal agitation and the resulting diffusion (Brownian movement) of proteins forbid localization in space on long timescale, since diffusion is a mixing process that will render every accessible position equiprobable. Inside eukaryotic cells, active mechanisms such as directed transport or sub-compartmentalization by internal membranes permit to counteract the uniforming effects of diffusion. It is however known since the 1952 seminal paper by Alan Turing [21] that subtle interactions between chemical reactions and diffusion can spontaneously lead to steady states with non-uniform spatial extension. This is also true for bacteria, as exemplified by the spatial oscillations in the minCDE system [22] or in the case of diffusion-trapping coupling [24]. Recently, the importance of precise sub-cellular localization of proteins within bacteria has become apparent [24,25,26]. In absence of a general cytoskeleton-based directed, active transport mechanism nor internal membranes, this would favor diffusion-reaction based localization within bacteria (see however [13,27,28]).

Specifically it is still unclear whether for single-cell organisms, preferential localization mechanism of damaged proteins is based on active directed transport or passive Brownian diffusion. In *S. cerevisiae*, initial reports incriminated a role for active directed transport (actin cytoskeleton) or sub-compartmentalization (membrane tethering) in the segregation of molecular damages (damaged proteins, episomal DNA) in the mother cell [14,29]. Yet, more recent reports contradict the need for directed transport, e.g. on the actin cable, and favor diffusion-based localization [16,30,31].

In *E. coli*, protein aggregates have consistently been reported to localize in the cell poles and in the middle of the cell [7,9,10]. The number of distinct aggregates per cell seems to depend on the cellular environment. In non-stressed conditions, at most one aggregate per cell is observed with rare cases (<4%) of two foci per cell detected [6]; under heat shock, most cells contain two or three aggregates [8,9]. In heat-shock conditions, Winkler *et al.* [9] concluded in favor of a Brownian passive motion of the protein aggregates. This study also pointed out that one simple possible passive aggregate localization mechanism may be based on spatially non-homogeneous macromolecular crowding. Indeed, in healthy cells, the bacterial chromosome spontaneously condensates [32] thus delineating a restricted sub-region of the cell called "nucleoid", where molecular crowding is much larger than in the rest of the cytoplasm [33]. Macromolecular crowding then alternates along the cell long axis between low intensity zones (cytosol) and large intensity ones (nucleoid). Monte-Carlo dynamics modeling suggests that such non-homogeneous spatial distribution of the molecular crowding may be sufficient to localize large proteins to the cell poles [34]. In line with this proposal are experimental reports that the observed aggregates preferentially localize in the nucleoid-free regions of the cell [7,9], i.e. precisely in the regions of alleged lower macromolecular crowding. In spite of these hints though, whether the transport of the aging-related protein aggregates in *E. coli* is of a directed active nature or purely passive Brownian origin remains elusive, since contradictory results indicate that this process would include ATP-dependent stages [9].

Here, our aim is to determine whether the movement of aging-related protein aggregates in *E. coli* is purely diffusive (Brownian) or includes some active process (ATP-dependent, directed transport or membrane tethering). To this aim, we devised an integrated approach combining time-lapse fluorescence microscopy of *E. coli* cells *in vivo*, open-source automated image analysis, and individual-based modeling. Our results strongly indicate that purely diffusive pattern of aggregates mobility combined with nucleoid occlusion underlie their accumulation in polar and mid-cell positions.

## RESULTS

**Trajectory analysis of single protein aggregates**
*In vivo* analysis of individual trajectories of proteins of interest (or aggregates thereof) is a powerful method to determine whether the movement of the target protein is of Brownian nature or additionally exhibits further ingredients (active directed transport, caging or corralling effects, transient trapping, anomalous sub-diffusion) [16,35-40]. Here, we focused on naturally forming protein aggregates tethered with the small heat-shock protein IbpA in *E. coli* whose spatio-temporal dynamics have been implicated in aging of the bacteria [7,13].

To characterize the motion of IbpA-tethered aggregates in single *E. coli* cells, we monitored intracellular trajectories of single foci of IbpA-YFP fusion proteins [7,41] in non-stressed conditions (37°C, in LB medium, see Materials and Methods). For the automatic quantification of the resulting time-lapse fluorescence microscopy movies, we developed dedicated image analysis and tracking software tools (see Materials and Methods). This software suite performs automatic segmentation and tracking of the cells (Figure 1A). Moreover, it automatically detects the fluorescence aggregates foci and monitors their movements relative to the cell in which they are located, with sub-pixel resolution.

*Localization of protein aggregates is non-homogeneous along the cells*
Detectable protein aggregates (in the form of localized fluorescence foci) were observed in half of the cells monitored (54%; $N_{cells}$ = 1625 recorded in 72 independent movies), in agreement with previous experimental reports [7,13]. No further foci were detected by doubling exposure time (see materials and methods). This suggests that smaller undetected aggregates that may exist either diffuse faster than the acquisition time and are therefore not recorded as "localized", or alternatively, that they merge into bigger, detectable aggregates before full maturation of the fluorophore (≤ 7.5 min). [42]. Cells in non-stressed conditions tend to exhibit smaller copy numbers of distinct protein aggregates than in heat-shocked conditions (compare e.g. [7] with [9,10]). Accordingly, in our hands, nearly all the foci-containing cells (98.8%) displayed a single fluorescence focus within the imaging time, while the remaining cells had at most two foci. The distribution of the (initial) spatial localization of the aggregates is displayed in Figure 1. As a convention, we denote the long axis of the bacteria cell as its $x$-axis and the short one as its $y$-axis (Figure 1B). In this figure, we express the aggregate position relative to the cell center of mass, and rescale to $[-1, +1]^2$ range. Thus the two bacterial poles correspond to $x = -1$ and $x = +1$ in this relative scale.

The histogram of the location of the aggregate at the starting point of each trajectory is shown in Figure 1C. The distribution is highly non-homogeneous, with most of the aggregates predominantly localized at one of the two poles, and the others mainly around the middle of the cell. This distribution is similar to the results obtained in [7] (Fig 2-E-F in [7]), except that here, since we do not differentiate between old and new poles, the amplitude of the polar modes in Figure 1C are roughly symmetrical. Similar distributions were also obtained in heat-shock conditions [9,10].

The marked localization of the aggregates suggests they might be tethered to the membrane at the poles (and center) at some point of the aggregation process, thus restricting their motion. Figure 1D shows the location of the polar aggregates at first detection (both poles were pooled). Because these experimental results are two-dimensional projections of three-dimensional positions, one cannot directly determine whether the aggregates are bound to the cell membrane or spread in the three dimensional cytoplasmic bulk. To this aim, we generated $10^4$ aggregate positions (uniformly) at random in a volume of the same shape and dimensions than the cell pole. Fig. 1E shows the two-dimensional projection of these positions when the proteins were randomly located in the three-dimensional bulk whereas Fig. 1F shows the two-dimensional projections when the proteins were randomly located on the cytoplasmic membrane enclosing the bulk.

To quantify these plots, we analyzed the local density of protein positions in the two-dimensional projections. Assuming the 2d projection of the pole is a semi-ellipse of radii $a_x$ and $a_y$ (green dashed shapes in Fig. 1 D-G), its area is $1/2\pi a_x a_y$. The area of the elementary semi-elliptic crescent $D_s$ (gray in Fig. 1G) delimited by the semi-ellipse of radii $sa_x$ and $sa_y$ (0<$s$<1) and that of radii $(s+ds)a_x$ and $(s+ds)a_y$ is thus $A(D_s)= \pi a_x a_y\, ds(s+ds/2)$. Varying $s$ between 0 and 1, we counted the number of aggregate positions $n_s$ found within the semi-elliptic disk $D_s$ and computed the corresponding density (correlation function) $\rho(s)=n_s/ A(D_s)$. Therefore, when $s$ approaches 1, $\rho(s)$ represents the local density of aggregate positions close to the external boundary of the pole (green dashed shapes in Fig. 1 D-F) whereas for small $s$ values, $\rho(s)$ gives the local density of aggregate positions close to center of the pole (i.e. the center of the semi-ellipse). To validate the approach, we generated random aggregate positions directly in two dimensions (uniform distribution inside the 2D semi-ellipse) and checked that the density $\rho(s)$ exhibits a constant value with $s$ (Fig. 1H, dotted black line).

When $\rho(s)$ was used to quantify the data of Fig. 1E (aggregates in the bulk), the density was roughly constant up to $s\sim0.3$ then decayed smoothly for larger values (Fig. 1H, blue line), in agreement with the slight decay of the aggregate density close to the semi-ellipse boundary that is already visible in Fig. 1E. By contrast, with the data shown in Fig. 1F (aggregates in the membrane around the bulk), $\rho(s)$ increased first slowly with increasing $s$, then very abruptly close to 1 (Fig. 1H, red line), in agreement with the accumulation of aggregate locations close to the semi-ellipse boundary that is already visible in Fig. 1F. Therefore, these simulated data show that the distortions due to the projection in two dimensions are expected to manifest as strongly increasing $\rho(s)$ values close to s=1 if the aggregates are tethered to the membrane versus smoothly decreasing $\rho(s)$ if the aggregates are randomly spread in the 3d bulk. Finally, we used this approach to quantify the experimental data described above (Fig. 1I). The local density $\rho(s)$ exhibits a non-monotonous behavior. Close to the pole boundary, $\rho(s)$ clearly shows an almost linear decay, which is very similar to the behavior observed at large $s$ for bulk simulations (Fig.1H). For small $s$ values however, $\rho(s)$ increases with $s$, thus indicating a phenomenon that hinders aggregate location close to the cell center. To conclude, these results plead in favor of a 3d bulk distribution of the aggregates in the poles, thus rejecting the hypothesis that they would be tethered or attached at the cell membrane.

*Polar protein aggregates exhibit Brownian motion*

We next quantified the displacements of the polar aggregates. To this end we used two temporal regimes in our time-lapse fluorescence microscopy. Low sampling frequency (LF; 0.33 Hz) over long time scale (LT; 5 minutes) was used to assure that we monitor entirely the range of displacements (n=1149), whereas high frequency sampling (HF; 1.67 Hz) was used to increase the temporal resolution of the linear initial regimes observed in mean-squared displacements. These conditions were optimized considering the trade-off between extensive exposure leading to bleaching of the foci and satisfactory temporal resolution.

The resulting mean displacements along the *x* or *y*-axis are shown in Figure S1. For both sampling frequencies, the mean displacement along the *y*-axis is roughly stationary and fluctuates around a close-to-zero value. In order to analyze the fluctuations, we rescaled the mean displacement along the *y*-axis so that it fluctuates around exactly zero.

The mean displacement along the *x*-axis displays a close-to-linear increase with time (Figure S1) with a slope of around 0.05 µm/min, a value that corresponds to the linear approximation of the cell elongation rate under our experimental conditions (doubling time around 30 minutes, during which the cell half-length grows from ≈ 1.0-2.0 to ≈ 2.0-3.5 µm). Therefore, the raw movement of the aggregates along the *x*-axis is dominated by ballistic displacement toward the pole under the effect of cell elongation. We corrected for this passive transport by subtracting the increase rate of the cell elongation.

The corrected mean displacements (Figure 2A) are stationary and slightly fluctuate around zero, for both HF and LF trajectories. This is a typical characteristic of Brownian motion ('random walk'). To confirm this hypothesis one has to study higher-order moments of the displacement, in particular the second one. The resulting mean-squared displacement $<(u(t)-u(0))^2>$ is displayed in Figure 2B. The LF and HF data here again are in very good agreement, with the HF data nicely aligned on the LF ones. This agreement is an important test of the coherence and quality of our measurement and analysis methodology. The inset of the figure shows a magnification of the HF data until time *t* = 30 sec. For the first 10 to 15 seconds, the HF data exhibits a clear linear behavior. As expected from an unbounded Brownian motion the same slope was observed for both the *x*- and *y*-axis. The non-zero intercept with the *y*-axis is typically due to the noise in the experimental determination of the aggregate position [43]. Such a linear dependence of the mean-squared displacement (MSD) is a further indication that the movement of the aggregates is Brownian diffusion, as one expects $<(u(t)-u(0))^2>=2D_u t$ in the case of a random walk (where $D_x$ or $D_y$ are the diffusion constant in the *x*- or *y*-direction, respectively). Using the first 15 seconds of the HF data, our estimates yield $D_x \approx 5.1\times10^{-4}$ µm²/s and $D_y \approx 4.0\times10^{-4}$ µm²/s. Note that these values are at best rough estimates since the data are averaged over aggregates of very variable initial sizes (whose mobility is expected to vary accordingly; see below). Nevertheless, the fact that the values for the *x*- and *y*-axes are similar is another indication of the isotropy of the Brownian motion that seems to govern the movement of the aggregates. These values are compatible with previous experimental reports of the diffusion constants of large multi-protein assemblies in bacteria, such the origin of replication in *E. coli* (around $10^{-4}$ µm$^2$/s [40]) but are significantly smaller than the values reported for single fluorescent proteins such as mEos2 or GFP (1 to 10 µm$^2$/s [39,44]).

Altogether the analysis of the first 15 seconds of the HF data pleads in favor of the hypothesis that the aggregates' motion is due to diffusion, thus excluding directed transport due to some active mechanisms. In order to further quantify the diffusive character of the aggregates at hand, we divided the LF data into 5 classes of increasing initial median fluorescence (table 1), so as to average data over aggregates of more homogeneous size. As depicted in Fig. 3A & B, the initial slope of the MSD vs. Time curves increases when the average intensity in the class decreases. Assuming that the initial median fluorescence is proportional to the initial size of the aggregate, this result further supports a diffusive behavior, for which the diffusion constant is expected to decrease as the molecular size increases. Plotting the data in log-log scale (Fig. 3C & D) evidences initial slopes (i.e. exponents of a possible power law) of 1.04+/-0.14 and 0.83+/-0.08 (on *y* and *x* direction respectively; mean slope, excluding the highest fluorescent class, see discussion). Note that the data for the y-direction are expected to be less noisy because they were not subjected to correction for cell growth, unlike the data for the x-direction. Moreover, in the x-direction, the aggregate movements appear to be restricted by a "soft" boundary (i.e. not a membrane, see below), which is expected to hinder interpretations of the MSD curves. Even so, the value of the exponent in the *x*-direction (0.83) is not significantly smaller than 1.0 (one-tailed t-test, 0.01 significance level). Therefore, we conclude from this data that the MSD at short times (before saturation) evolves linearly with time (MSD ~ $t^\alpha$ with α = 1), as expected from a Fick-like normal diffusion. This is in contrast with recent reports where anomalous diffusion was recorded (α in [0.40,0.75]) for RNA-protein assemblies [45,46] and further supports our conclusion favouring passive pure diffusion mechanism of protein aggregates in *E. coli.* In our results, anomalous diffusion can be excluded except for the largest aggregate class (black circles in fig. 3A-D); however, the movement amplitude of these very large aggregates is too low to allow precise quantification. We further controlled for growth rate and cell-length effects on the diffusion pattern observed. Among the different cells we imaged, the variations in cell growth were very small, whereas cell length varied appreciably. However, the initial slopes showed no significant differences when the data were clustered into sub-classes of cell lengths (Figure S2).

*Protein aggregates exhibit passive confinement within bacterial polar*

*region*

Following the short, approximately 15 seconds linear regime, the MSD change with time decelerates (inset, Fig. 2B) and reaches full saturation after about 40 seconds (LF data, Fig. 2B). MSD Saturation occurs in both x- and y-data at values of 0.030 and 0.040 µm², respectively. This corresponds to a restriction of the movements of the aggregates in a sub-region of the bacterial interior, with characteristic size ≈ 400 nm. To understand these results, one has to take into account the size variability of the monitored aggregates. Indeed, when considering the 5 sub-classes of aggregates (Table 1), the observed saturation levels are not constant but clearly decrease when the initial total fluorescence increases. This suggests that the size of the large aggregates is of the same order than the cell dimension, so that the intracellular space available for an aggregate of size $r$ is in fact $L-r << L$ (where $L$ is the cell size). In turn, this enables us to estimate the actual size of the foci, independent of the microscope resolution.

To this end, we used the data in Figure 3A and C to obtain reliable estimates of the sizes of the aggregates and that of the intracellular subregion in which they are confined. We used an automatized procedure based on parameter optimization by an evolutionary strategy of the parameters of an individual-based simulation for constrained diffusion (see Materials and Methods for full description). In short, our strategy can be seen as an automatized fit of the values of 12 parameters: the dimensions $L_X$ and $L_Y = L_Z$ of the intracellular subregion in which the aggregates are confined and the average radius $r_i$ and diffusion constant $D_i$ of the aggregates belonging to class $i$ ($i$ =1..5). We start by setting these parameters to initial guess values. We then simulate the confined diffusion of 5,000 spherical molecules per class that are endowed with the corresponding values of $r_i$ and $D_i$ ($i$ =1..5) and diffuse in a spatial domain which size is given by the initial guess values of $L_X$ and $L_Y = L_Z$. The values of the MSD (averaged over each class) in the $x$ and $y$ directions are then sampled during the simulation at the same frequency (0.33 Hz) as in the LF data of Figure 3A and C. We then compute the distance between the resulting curves for the simulated data and those for the experimental LF data. This distance represents the least-square error between the MSD predicted by the simulations using the initial values of the parameters. To minimize this error automatically, we used an evolutionary strategy called CMA-ES (see [47] and Material and Methods). CMA-ES automatically finds the values of the parameters that yield the smallest distance between experimental and simulation data. Note that because the correction procedure (that removes the effect of cell elongation) for the $x$-data is strong (i.e. the aggregate motion in this direction is originally dominated by cell elongation), it is difficult to apply this fitting procedure on the experimental MSD data for the $x$-and y-axis simultaneously. This obliged us to fit the parameters separately on the *x* and *y*-data. The numerical values indicated in table 1 represent the average of these two fits (the "±" values indicate the total variation range).

The class-averaged MSD corresponding to the optimized parameters are shown in Figure 3A&C as lines and are in agreement with the experimental data. The best-fit values for the cell dimension parameters are $L_X$ = 650 nm and $L_Y$ = $L_Z$ = 750 nm. Considering our pixel size (64 nm), we conclude that the value of the space available to the aggregate in the *y* and *z* directions is close to the whole space available in this direction (around 900 nm). Therefore the aggregate motion in the *y* and *z* directions seems to be constrained only by the inner cell membrane. In strong contrast, while the total intracellular space in the *x*-direction ranges from 2 to 5 µm (depending on the cell elongation), our fitting procedure indicates that the aggregates do not move beyond 650 nm in this direction. As they are initially located in the cell pole, this shows that the aggregates remain in the pole, where a passive mechanism hinders their free movement and keep them from leaving a subregion spanning ≈ 1/4th to 1/9th of the total cell length. This is coherent with the positioning of the nucleoids, hindering the diffusion of the aggregates (*see below)*.

The experimental relation between the aggregate radius and diffusion constant, as determined by the fitting procedure above (listed in table 1), shows a good fit to a Stokes-Einstein relation: $D(r)=C_0/r$ where one expects $C_0=kT/(6\pi\eta)$ (describing simple sphere diffusion in a liquid of viscosity $\eta$) (Fig. 3E). We obtain a remarkable fit to the experimental data with $C_0$ = 47230 nm$^3$/s. Note that a power-law fitting $D(r)=C_0/r^b$ gives *b*=0.89+/-0.14 (thus that does not exclude *b*=1.0) with a similar quality of fit (chi-square values). In the inset of Figure 3E, the experimentally determined values of *D* are plotted as a function of 1/*r*. The relation is linear except perhaps for very large values of *r*, thus emphasizing the very good agreement with the Stokes-Einstein law.

These results confirm that some passive mechanism hinders the free movement of the aggregates along the long axis. They also show that the aggregates follow the Stokes-Einstein law even for large sizes since the diffusion constant of the aggregates simply decays as the inverse of its radius.

## Simulation of aggregation, Brownian diffusion, and molecular crowding reconstitutes the observed patterns

If the spatial dynamics of protein aggregates in *E. coli* is indeed based exclusively on Brownian motion and aggregation, we should be able to reproduce by computer simulations the observed spatial patterns of the aggregates described here and in previous experimental reports [7,9,10]. In particular, given the relative simplicity of these two ingredients (diffusion + aggregation), individual-based modeling is a very convenient tool in this framework. Here, we used an individual-based model in which proteins diffuse via lattice-free random walks in a 3D domain that has the size of a typical *E. coli* cell (see Material and Methods and Figure 4A). When two molecules meet along their respective random walks, they form a single aggregate with probability $p_{ag}$. The size of the resulting aggregate is updated according to the size of its constituting molecules and its diffusion constant is updated as a function of the aggregate size using a simple Stokes-Einstein law. In turn, such aggregates can combine to form bigger aggregates.

The above experimental finding suggests that the aggregate's Brownian motion is restricted in the *x*-direction over a distance that is comprised between $1/4^{th}$ to $1/9^{th}$ of the cell long axis. These numbers are in agreement with the hypothesis that the physical structures that hinders aggregate diffusion are the nucleoids, i.e. the subregion of the bacterial cytoplasm where the chromosome condensates, thus increasing molecular crowding and diffusional hindrance [33]. In a typical *E. coli* culture in exponential phase, most cells contain two nucleoids [48] and the free space between each nucleoid and the closest pole end is around 500-600 nm (deduced from [49]). We thus tested the hypothesis (already pointed out in [7] or [10]) that the restriction to the free movement of the obstacles along the long axis of the bacteria is caused by an increased molecular crowding in the nucleoids. To simulate this increased molecular crowding in our individual-based model, we added immobile non-reactive obstacles to represent the densely packed nucleoids (Fig. 4A, see Materials and Methods).

Each simulation is initiated by positioning $N_p$ single proteins at random inside the cells. The simulation then proceeds by moving the proteins via random walks and letting them aggregate (with resulting update of the aggregate radius and diffusion constant) when they encounter. The initial number of proteins $N_p$ was varied in order to account for different experimental conditions. We used $N_p$=100 to emulate experiments in non-stressed conditions (as in this work) where one does not expect the presence of large numbers of proteins in the aggregates. For experiments where aggregation is intensified using for example heat shock treatment [9,10], one generally expects to recover more proteins in the aggregates. In such conditions, the number of proteins in the aggregates was evaluated to be between 2,000 and roughly 20,000 molecules [10]. Here, to account for these data, we used at most a total of $N_p$

= 7,000 molecules in the simulations.

In the experiments, protein aggregates are detected by fluorescence microscopy when their size is large enough; though we have currently no means to quantify this threshold size, we estimate our detection limit to contain about 10-50 YFP copies, based on the detection of slow mobile particles [50]. Therefore, in the simulations, we had to arbitrarily fix a threshold for the number of proteins per aggregate above which the aggregate is considered large enough to be detected. In non-stressed conditions ($N_p$ = 100), we used a detection threshold of 30 proteins per aggregate, whereas with $N_p$ = 7,000 total molecules per cell, we varied the detection threshold between 30 and 1750 monomeric proteins per aggregates.

We first focused on the distribution of the position of the aggregates at their first detection along the bacterial long axis. The corresponding distribution is shown in Figure 4B. In non-stressed conditions ($N_p$ = 100), the distribution of the aggregate position is qualitatively very similar to our experimental results (see Figure 1C): most of the aggregates are located in the poles, the rest being mainly located in the cell center. This result is largely robust to the value of the detection threshold of the aggregates. For very large values of the detection threshold, the distributions of the aggregate location tend to decay inside the nucleoids and to increase in the poles. However, outside such extreme ranges, the spatial distribution of the aggregates is qualitatively robust to the detection threshold. In Fig. 4B, for instance, the full lines correspond to aggregate detection thresholds ranging from to 5 to 50 proteins per aggregate. The resulting spatial distributions are however similar and match all well with the experimental result (dashed line histogram). This robustness is also observed when the initial locations of the proteins are taken inside or immediately around the nucleoids (Figure S3A & C). Importantly, in agreement with experimental reports, the nucleoids in the simulations are not fully impermeable to the aggregates, since the probability to observe large aggregates in the nucleoids is not null. Rather, the presence of the largest molecular crowding in the nucleoids reduces the probability that large aggregates form in the nucleoids.

In addition, starting the simulation with a non-homogeneous distribution of proteins, such as confinement to the nucleoid area or just englobing it, does not change the aggregate number per cell and distribution outcome, though a short delay in the accumulation of aggregates in observed (Figure S3, compare e.g. B & D with Fig. 4C & E, respectively). This is dictated by the crowded nucleoid model we adopted with realistic mesh size that does not prohibit diffusion of monomeric proteins within the nucleoid crowded area.

Another intriguing experimental result is that the number of aggregates simultaneously present in a cell depends on the experimental conditions. In non-stressed conditions only 1.2% of the observed cells presented more than a single aggregate within the observation time [7] whereas in heat-shocked conditions a majority of the cells present two distinct aggregates, and the rest is roughly equally distributed between one and three protein aggregates per cell. These observations are faithfully reproduced by diffusion-aggregation mechanisms. Indeed, in our simulations emulating non-stressed conditions (100 proteins), a vast majority of the simulations develop a unique detectable aggregate within the simulation time (Fig. 4C). Despite the increase with time of the probability for a simulation to feature two detectable aggregates per cell, the probability to observe only one aggregate remains largely higher, even at the end of the simulations.

In simulated heat-shock conditions (see Figure 4D, total of 7,000 proteins) with a very low aggregate detection threshold, virtually all simulations display at least 4 distinct detectable protein aggregates per cell from the very beginning of the simulation. These results are comparable with the experimental data obtained in [7], where the addition of streptomycin was shown to strongly increase the size and number of detected aggregates per cell (to 5 or more). When we increased the aggregate detection threshold (illustrated by a detection threshold of 1750 aggregates in Fig 4E) the simulations showed very different behavior. For short simulation times, we mainly observed cells with a unique detected aggregate. As simulation time increases, the probability to detect a unique aggregate decays in favor of the probability of detect 2 and 3 aggregates simultaneously in the cells. Eventually, most of the simulations (around 60%) display two aggregates, and the others are roughly evenly shared between 1 and 3 detected aggregates per cell, in agreement with the experimental results reported in [10] that employ heat-shock triggered aggregation.

These simulations predict that the number of detected aggregates in the cell is crucially dependent on two main factors: the aggregate detection threshold and the total number of aggregation-prone proteins. Therefore they suggest that the discrepancy observed concerning the number of aggregates per cell between non-stressed and heat shock conditions is due to the larger quantity of aggregate-prone proteins resulting from the heat shock. Taken together, our results show that the three basic ingredients we considered in our simulations (passive Brownian motion, aggregation, increased molecular crowding in the nucleoids) are sufficient to reproduce several experimental observations on the spatial distribution and number of protein aggregates in *E. coli.* Therefore, they confirm the conclusion drawn from our experimental results above that the movement of the chaperone protein aggregates in *E. coli* is driven by passive diffusion (Brownian motion). They moreover indicate that the observed non-homogeneous spatial distribution is not due to active or directed aggregate movement but is a mere result of the interplay between Brownian diffusion and molecular crowding.

# DISCUSSION

Our objective in this work was to decipher the mechanisms by which protein aggregates in *E. coli* localize to specific intracellular regions, *i.e.*, cellular poles.

Using single-particle tracking of protein aggregates marked with the small heat shock chaperone IbpA (inclusion-proteins Binding Protein A) translationally-fused to the yellow fluorescence protein (YFP), our results indicate that protein aggregate movements are purely diffusive, with coefficient constants of the order of 500 nm$^2$/s, depending on their size. Noteworthy, recent quantification of the movements and polar accumulation in the poles of MS2 multimeric RNA-protein complexes and fluorescently-labelled chromosomal loci concluded a high degree of anomalous diffusion, as reflected by slopes of 0.4-0.75 in log-log plots of time-MSD relationships [45,46]. This suggests that unlike pure protein aggregates, these complexes have further significant interactions with cellular components.

Applying evolutionary strategy for parameters estimation under the hypothesis of confined diffusion, we used our experimental data to estimate the average size and diffusion constant of the aggregates and the distances over which their movement is confined. As expected, the aggregate diffusion constant decreases with increasing aggregate sizes, but, more surprisingly, we find that the relation between the aggregate diffusion constant and their size is in very good agreement with the Stokes-Einstein law, thus strengthening the demonstration of pure Brownian motion. The agreement with the Stokes-Einstein law, that predicts a decay of the diffusion constant as the inverse of the radius, *D*~1/*r*, was found valid for all the estimated aggregate radii, even as large as 250-270 nm.

Recent experimental tests of the validity of this law in *E. coli* were more ambiguous. Kumar and coworkers [51] quantified the diffusion of a series of 30–250 kDa fusion proteins (some of which contained native cytoplasmic *E. coli* proteins) in *E. coli* cytoplasm and found very strong deviation from the Stokes-Einstein law -even for small proteins- with very sharp decay of the diffusion constant *D*~1/*r*$^6$. However, using GFP multimers of increasing sizes, Nenninger *et al.* [52] found very good agreement with Stokes-Einstein law from 20 to 110 kDa, i.e. up to tetramers, while the diffusion constant for pentamers (138 kDa) was found smaller than Stokes-Einstein prediction. Moreover, deviations from the Stokes-Einstein law was suggested an indication of specific interactions of the diffusing protein with other cell components. A tentative interpretation of our observation that even large cytoplasmic protein aggregates in *E. coli* do follow Stokes-Einstein law, would be that these aggregates actually have limited interactions with other cell components. This hypothesis would match very well with the putative protective function of the aggregates as scavengers of harmful misfolded proteins, allowing their retention within large, stable objects [1].

A second major finding of our study is the demonstration that the Brownian motion of the aggregates is restricted by the cell membrane in the section plane of the cell, while, along the cell long axis, the aggregates are confined to a region that roughly corresponds to the nucleoid-free space in the pole, thus confirming the importance of hindered diffusion in the nucleoids. In further support to this hypothesis, we used 3D individual-based modeling to show that these three ingredients are sufficient to explain the most salient experimental observations. Our simulations exhibit spatial distributions of the aggregates that are similar to those observed in non-stressed as well as heat-shock conditions. They also explain the differences in the number of distinct aggregates per cell as a mere difference in the total number of aggregation-prone (misfolded) proteins. Therefore, our results strongly support the hypothesis that the localization of aging-related protein aggregates in the center and poles of *E. coli* is due to the coupling of passive diffusion-aggregation with the spatially non-homogeneous macromolecular crowding resulting from the localization of the nucleoid(s).

Our computational approach can be further extended to address asymmetric division of cellular components in dividing cells. Due to computation time limitations inherent to individual-based models, a valid approach to pursue would be to derive a mean-field model of the diffusion-coagulation process, using e.g. integro partial differential equations with position-dependent properties for the diffusion constant or operator (Laplacian) combined with a coagulation operator [53]. This approach would allow to reach simulated times large enough to account for several cell generations and focus on the location of the larger aggregates along the lineage.

As a whole, our results emphasize the importance for diffusion-based protein localization of the "soft" intracellular structuring of *E. coli* along the large axis due to increased macromolecular crowding in the nucleoids. In addition to this implication in the localization of aging-related protein aggregates, the structuring effect of the nucleoids has very recently been evidenced in the accurate and robust positioning of the divisome proteins (that mediate bacterial cytokinesis) [54] or the non-homogeneous spatial distribution of the transcription factor LacI [55].

Wild-type *E. coli* cells (*e.g.*, without any expression of fluorescent proteins) exhibit qualitatively and quantitatively the same ageing phenotype in terms of gradual fitness as those expressing fluorescent markers (as the IbpA-YFP fusion) [7]. Yet, in previous studies, limited to very few ageing generations (typically less than 8), visible aggregates in wild-type ageing cells were seldomly detected, probably because of the phase contrast detection threshold (typically 500nm) [7]. To ensure that polar accumulation of aggregates is indeed a phenotype of wild-type cells, we used the recently developed 'mother machine' microfluidics system [12] to grow ageing wild-type *E. coli* through >150 generations. Under these conditions, many of the ageing cells indeed accumulate clearly visible aggregates (Figure S4), pointing to the validity of our approach to use the IbpA-yfp system for better detection [7]. The mechanism described here for IbpA-yfp tethered aggregates can be generalized as ample evidence exist for polar localization of aggregates resulting from heterologous over-expression of proteins, streptomycin treatment [7 and ref therein], large protein assemblies of fluorescently-labelled protein fusions (due to avidity of low multimerization propensity of some fluorescent proteins and independent of the diffusive positioning of the native proteins studied [13]), large RNA-protein assemblies [45,46]. In all cases, given the non-specific nature of hydrophobic interactions governing aggregate assembly, it is unsurprising that co-localization may occur amongst different aggregated polypeptides and chaperones, based on the common diffusive mechanism of polar accumulation described here. Moreover, our recent work demonstrates that large engineered RNA assemblies accumulate as well in the cells' poles [56, (electron microscope images therein)]. Therefore, the polar localization pattern of low diffusive elements in bacteria is not limited to large purely protein assemblies. We propose that it might be a more general process concerning other cell constituents, such as nucleic acids.

# MATERIALS AND METHODS

**Bacterial strains.**

The sequenced wild-type *E. coli* strain, MG1655 [57] was modified to express an improved version of the YFP fluorescent protein fused to the C terminus of IbpA [35] under the control of the endogenous chromosomal ibpA promoter resulting in the MGAY strain. *E. coli* strains were grown in Luria-Bertani (LB) broth medium half salt at 37°C. For more information about the cloning, see [7], *S.I*.

**Fluorescence time-lapse microscopy setup.**

After an overnight growth at 37°C, MGAY cultures were diluted 200 times. When the cells reached an absorbance 0.2 (600 nm), they were placed on microscope slide that was layered with a 1.5% agarose pad containing LB half salt medium. The agarose pad was covered with a cover-slide, the boarder of which was then sealed with nail polish oil.

Cells were let to recover for 1 hour before observation using Nikon automated microscope (ECLIPSE Ti, Nikon INTENSILIGHT C-HGFIE, 100x objective) and the Metamorph software (Molecular Devices, Roper Scientific), at 37°C. Phase contrast and fluorescence images (25% lamp energy, 1 second illumination LF movies and 600 milliseconds for HF movies) were sampled at two different time-scales. For low-frequency (LF) movies, images were taken every 3 seconds for a total of 5 minutes, while for high-frequency (HF) movies, fluorescence images were taken about very 0.60 seconds for a total of 2 min (and phase contrast images were sampled about every 7 fluorescence images). Fluorescence excitation light energy level used here is 5-fold higher than previously described [7] to allow proportional decrease of exposure time, enabling a higher temporal resolution. Under these conditions, doubling the exposure time did not result in further detection of fluorescent foci yet resulted in accelerated bleaching that prevented consecutive time lapse imaging of the observed foci.

**Image analysis and aggregate tracking.**

Phase contrast images were analyzed by customized software "Cellst" [58] for cell segmentation and single cell lineage reconstruction. Phase contrast images were denoised using the flatten background filter of Metamorph software for long movies or a mixed denoising algorithm [58] for fast movies. The mixed denoising algorithm combines two famous image denoising methods: NL-means denoising [59], which is patch-based and Total Variation denoising [60,61], which is used as regularization. The Cellst software was used to automatically segment the cells on most of the images, albeit when necessary, manual corrections were applied. At the end of the whole segmentation and tracking process, Cellst also calculates the lineage of every cell in the movie.

The fluorescent protein aggregates were detected by another customized software. Detection of the maximal intensity pixel of each spot was realized using the a contrario methodology based on a circular patch model with a central zone of detection and an external zone of context. The patch radius was then optimized to optimally match the spot. The energy of each spot was computed in the following way: the total image was modeled as a sum of a constant background and 2D circular Gaussian curves, centered on the maximal intensity pixel of the detected spots with a deviation of 3 pixels. The quadratic minimum deviation between the image and the model enabled to calculate the Gaussian coefficients. These coefficients were considered the energy values of each spot. The coordinates of the detected spots were then refined to subpixel resolution. This was achieved by computing a weighted average of the coordinates of the pixels in a circular neighborhood of the detected spot. The weights were given by the intensity values of the pixels to which the local background is subtracted. The local background was then computed as the median value of the pixel intensity in the neighborhood. Only pixels having intensity bigger than the median value were considered in the weighted average. This algorithm has been tested on both synthetic and real image and it shown a precision of 1/10 of a pixel on very poor contrasted spots.

After detection and localization, the movements of the fluorescent aggregates were tracked and quantified by a third customized software named "aggtracker" based on the cell lineage and the detected spots. The algorithm uses the lineage and cell information to ensure that an aggregate is consistently tracked through points with points that are inside the same cell.

The output of this software are time-series for the coordinates x and y (in pixels) of each fluorescence spot as well as the affiliation of the spot to the cell it is in. The last step consisted in the projection of the coordinates of the fluorescence spot from their initial absolute values in the image (in pixels) to their value along the long and short axes of the 2d image of the cell. To this aim, we used active skeletons. A skeleton represents an object by a median line (the center line in the case of a tubular bacteria). Here we used one active skeleton for the long axis of the cell image and a second one for its short axis. Active skeletons were adapted to bacteria in order to optimize the position of the skeleton in the image of the cell. The coordinates of the fluorescence spots were then expressed as the coordinate of the center of the fluorescence spot in the basis composed of the two active skeletons that localize the cell long and short axes. We exploited the simple shape of the skeleton to estimate the total cell width and length as that of the respective skeleton. As a convention, we refer below to the aggregate coordinate along the long axis as the x-coordinate and that along the short axis as the y-coordinate.

In order to improve precision, aggregate trajectories made of less than 10 successive images in the movies were not further taken into account. In total, we obtained 1644 aggregate trajectories.

**Individual-based modeling of protein aggregation.**

To simulate the diffusion and aggregation process of proteins in a single cell, we used a 3d individual-based lattice-free model. Each protein $p$ was explicitly modeled as a sphere of radius $r_p$ centered at coordinates ($x_p$, $y_p$, $z_p$) in the 3d intracellular space of the cell. We simulated protein diffusion in the cell and aggregation as they encounter using as realistic conditions as possible. In particular, the radius and diffusion coefficient of the protein aggregates explicitly increased as they grow. Moreover, we explicitly modeled the larger molecular crowding in the nucleoids. Details of the simulations are as follows.

The bacterial cell was simulated as a 3D square cylinder with width and depth 1.0 µm [62] and length 4.0 µm (chosen to correspond to a bacterial cell just before division) and reflective boundaries. Note that we also ran simulations with more realistic cell shapes (i.e. spherical caps at cell ends) and did not find significant differences compared to square cylinders (except for the much higher computation cost with spherical caps).

Within each cells, we also explicitly modeled the larger molecular crowding found in the nucleoids. Indeed, in healthy cells, the bacterial chromosome condensates into a restricted sub-region of the cell called "nucleoid", where molecular crowding is much larger than in the rest of the cytoplasm [33]. To model this increased molecular crowding in the nucleoids, we placed at random (with uniform probability) 50,000 bulky immobile, impenetrable and unreactive obstacles (radius 10 nm) in the region of the cell where a nucleoid is expected. Because cell cycle and DNA replication in *E. coli* are not synchronized, roughly 75% of the cells in exponential phase contain two nucleoids [48]. We thus explicitly positioned two nucleoids within the cell. The location and size of the two nucleoids were estimated from DAPI-stained inverted phase contrast images of the nucleoids found in [49]. Both nucleoids were 3d square cylinders of length 1220 nm (along the cell long axis) and width and height 532 nm. Each nucleoid started at 540 nm from each cell pole and was centered on the cell long axis. The volume occupied by the two nuclei area thus formed is about 12%, which is consistent with literature [63,64].

Each simulation was initialized by positioning $N_p$ individual IbpA-YFP proteins (monomers) at non-overlapping randomly chosen (with uniform probability) locations in the free intracellular space of the cell (i.e. the whole interior of the cell minus the space occupied by the obstacles in the nucleoids). At each time step, each molecule is independently allowed to diffuse over a distance $d$ that depends on the protein diffusion constant $D_p$, according to $d = (6\ D_p\ \Delta t)^{1/2}$, where $\Delta t$ is the time step, in agreement with basic Brownian motion. Note that $D_p$ itself depends on the aggregate size $r_p$ (see below). The new position of the protein $(x',y',z')$ was then computed by drawing two random real numbers, $\theta$ and $c$, uniformly distributed in $[0, 2\pi]$ and $[-1,1]$, respectively, and spherical coordinates: $x'=x(t)+d\ \sin(\mathrm{acos}(c))\ \cos(\theta)$; $y'=y(t)+d\ \sin(\mathrm{acos}(c))\ \sin(\theta)$ and $z'=z(t)+d\ c$ where $(x(t),y(t),z(t))$ is the initial position of the protein. If the protein in this new position $(x',y',z')$ overlaps with any of the immobile obstacles (i.e. if there exists at least one obstacle such that the distance between the obstacle center and $(x',y',z')$ is smaller than the sum of their radii) the attempted movement is rejected $(x(t+\Delta t),y(t+\Delta t),z(t+\Delta t)) = (x(t),y(t),z(t))$. This classical approximation of the aggregate reflection by the static obstacles is not expected to change the simulation results significantly, but it drastically reduces the computation load. If no obstacle overlaps, the movement is accepted, i.e. $(x(t+\Delta t),y(t+\Delta t),z(t+\Delta t))= (x',y',z')$.

After each molecule has moved once, the algorithm searches for overlaps between proteins. Two proteins are overlapping whenever the distance between their centers is smaller than the sum of their radii. Each overlapping pair was allowed to aggregate with (uniform) probability $p_{ag}$ (irrespective of their size). In our simulations, $p_{ag}$ was varied between 0.1 and 1.0 (limited at the lower band by simulation time needed to score enough aggregation events). To model the aggregation from two overlapping proteins, we could not, for computation time reasons, keep track of the shape of the aggregates (i.e. the individual location of each protein in the aggregates). Instead, we used the simplifying hypothesis that all along the simulation, the aggregates maintain a spherical shape with constant internal density. It follows that the radius of an aggregate C, born out of the aggregation of two aggregates A and B of respective size $r_A$ and $r_B$ is $r_C = (r_A^3 + r_B^3)^{1/3}$. Upon aggregation, we thus remove the aggregates A and B from the cell, and add a new aggregate with size $r_C$, centered at the center of mass of the two former aggregates A and B.

Finally, to set the diffusion constant of the aggregates, we used the classical Stokes-Einstein relation for a Newtonian fluid, where the diffusion constant is inversely proportional to its radius. In our case, this leads to $D_p = D_0 r_0 / r_p$ where $r_0$ and $D_0$ are the radius and diffusion constant, respectively, of individual (monomeric) IbpA-YFP molecules. Note that this relation could be violated for large molecules in the cytoplasm of *E. coli* [52,53]. In a subset of simulations, we used power law relations, such as $D_p \propto r_p^{-6}$, as suggested in [51], without noticeable change in our results (except that the time needed to reach a given threshold aggregate size was increased). Note that aggregation was considered irreversible in our model (i.e. aggregates do never breakdown into smaller pieces). This is in agreement with our experimental observations, where we never measured decay of foci fluorescence.

The diffusion constant of the 26-kDa GFP (radius 2 nm) in *E. coli* cytoplasm is around 7.0 µm$^2$/s and that of the GFP-MBP fusion (72 kDa) around 2.5 µm$^2$/s [37]. Using this data and the Stokes-Einstein relation combined to our constant spherical hypothesis, led to estimates of the radius and diffusion coefficient of the individual (monomeric) 39 kDa IbpA-YFP fusion of $r_0$= 3 nm and $D_0$=4.4 µm$^2$/s.

The value of the time step $\Delta t$ has to be small enough so that proteins cannot jump over each other during a single time step, meaning that the distance diffused during a single time step is limited by $d < 4r_0$. Using the definition for $d$ above, one then has $\Delta t < 8/3 \, r_0^2/D_0 \approx 5$ µs. Here, we used $\Delta t$ = 1 µs yielding $d$= 5 nm for monomeric proteins.

Every simulation was run for a total of 2×10$^6$ time steps. The translation of this value into real time is hardly possible since we have no indication of the experimental value of the aggregation probability per encounter $p_{ag}$ (see above) even less so of its dependence on the aggregate size. A lower bound can be estimated to 2 seconds real time (for 2×10$^6$ time steps) if the aggregation is always diffusion limited (i.e. $p_{ag}$=1). On general grounds however, the experimental value of $p_{ag}$ can be expected to be smaller, so that the 2×10$^6$ simulation time steps would correspond to more than this 2 seconds real time minimal value. For the results to be statistically significant, we ran $n_{run}$ simulations for each parameter and condition, with different realization of the random processes (initial location, random choice of the positions or of the aggregation events) and averaged the results over these $n_{run}$ simulations. In the results presented here we used $n_{run}$ = 10$^3$.

**Fitting procedure for the aggregate radius, diffusion constant and cell dimensions**
The data from the LF movies were partitioned into 5 classes based on the aggregate fluorescence intensity at the beginning of the measured trajectory, yielding 5 pairs of experimental curves for the mean-squared displacement, $<(x_i^{exp}(t)-x_i^{exp}(0))^2>=f_x(t)$ and $<(y_i^{exp}(t)-y_i^{exp}(0))^2>=f_y(t)$ where $i$={1,…,5} indexes the intensity class. Corresponding theoretical values were obtained by individual-based simulations of confined random walks similar to those described above but modified as follows: the cells, of dimensions $LX$ (length), $LY = LZ = LYZ$ (height and width) were devoid of nucleoids or aggregation (aggregation probability $p_{ag}=0$) and we used $N=5,000$ IbpA-YFP proteins. Each 12-uplet of parameters $\{LX, LYZ, r_i, D_i\}$ yields two theoretical curves ($<(x_i^{the}(t)-x_i^{the}(0))^2>=g_x(t)$ and $<(y_i^{the}(t)-y_i^{the}(0))^2>=g_y(t)$). The aim of the fitting procedure is to minimize the distance between the experimental and theoretical curves, ie to minimize the cost function:

$$F = \sum_{i=1}^{5} \sum_{j=1}^{N} \left( \frac{\langle (x_i^{exp}(t) - x_i^{exp}(0))^2 \rangle - \langle (x_i^{the}(t) - x_i^{the}(0))^2 \rangle}{\langle (x_i^{the}(t) - x_i^{the}(0))^2 \rangle} \right)^2 +$$

$$\left( \frac{\langle (y_i^{exp}(t) - y_i^{exp}(0))^2 \rangle - \langle (y_i^{the}(t) - y_i^{the}(0))^2 \rangle}{\langle (y_i^{the}(t) - y_i^{the}(0))^2 \rangle} \right)^2$$

where the indices $j$ are over the $N$ time steps. The formulation of this cost function corresponds to the traditional least squares, so that the optimization procedure actually looks for best fits in the least-square sense (minimization of the squared residuals between the theoretical predictions and experimental observations). To minimize automatically the cost function $F$, thus adjusting the theoretical to the experimental curves, we used the C++ implementation of the state-of-the-art evolutionary strategy algorithm CMA-ES [39] with population size 12 and 400 generations.

# ACKNOWLEDGEMENTS

We thank N. Hansen, for providing the source code of CMA-ES for various programming languages (downloadable at http://www.lri.fr/~hansen/cmaes_inmatlab.html). We also thank the CNRS-IN2P3 Computing Center (cc.in2p3.fr) for providing computer resources.

Biophys Chem 73(1-2): 23-29.

# FIGURE LEGENDS

**Figure 1**: **Localization of the detected aggregates in the cells**
(*A*) In each image on the time-lapse fluorescence movies, the bacterial cells are automatically isolated (each individual cell is given a unique random color). The aggregates appearing during the movie are automatically detected and their trajectory within the cell quantified (internal trajectories). (*B*) By convention, we referred to the projection of the aggregate location on the long axis of the cell as the *x*-component and that along the short axis as the *y*-component. (*C*) Histogram of the *x*-component of the initial position of the trajectories (total of 1,644 trajectories). Since the cell length at the start of the trajectory is highly variable, the *x*-component was rescaled by division by the cell half-length. After this normalization, the cell poles are located at locations -1.0 and 1.0 respectively, for every trajectory. (*D*) Experimentally measured positions of the aggregates detected in the poles (both poles pooled, $n$ = 9,242 points). The green-dashed curves in (*D-F*) locate the 2d projection of the 3d semi-ellipsoid that was used to approximate the cell pole. (*E*) Synthetic data for bulk positions: 10,000 3d positions were drawn uniformly at random in the 3d semi-ellipsoid pole. The figure shows the corresponding 2d projections. (*F*) Synthetic data of membranary positions: 10,000 3d positions were drawn uniformly at random in the external boundary (membrane) of the 3d semi-ellipsoid pole. The figure shows the corresponding 2d projections. (*G*) To quantify figures D-F, the correlation function $\rho(s)$ was computed as the density of positions located within crescent $D(s)$ (gray). See text for more detail. (*H-I*) Local density of aggregate positions $\rho(s)$ in the synthetic (*H*) and experimental (*I*) data shown in *E* (bulk, blue), *F* (membranary, red) and *D* (experimental, orange). The dashed black line shows the local density computed for 10,000 synthetic *2*d positions that were drawn uniformly at random in the 2d semi-ellipse resulting from the 2d projection of the 3d pole ellipsoid (green dashed curve in *D-F*).

**Figure 2: Single-aggregate tracking analysis inside *E. coli* cells.**
Coordinates along the *x* and *y*-axis are shown in red and black, respectively. Low frequency sampling trajectories (LF) are displayed using full lines and high frequency ones (HF) using open symbols. Light red and black swaths indicate + and - 95% confidence intervals for the *x*- and *y*-axis data, respectively (for clarity, - and + intervals for the x- and y-axis data, respectively, are omitted) (*A*) Corrected mean displacement $<(u(t)-u(0)-u_c(t))^2>$ where $u_c(t)$ is the applied correction. For the *y*-component, the correction is the time-average of the *y*-coordinate. For the *x*-component, the applied correction is cell growth : $u_x=(L(t)-L(0))\Delta t$ where $L(t)$ is the cell half-length at time *t* and $\Delta t$ is the time interval between two consecutive images. (*B*) Corresponding mean squared displacements MSD$c$ = $<(u(t)-u(0)-u_c(t))^2>$. The inset shows a magnification of the HF results and their close-to-linear behavior for the first 10 - 15 seconds (dashed line).

**Figure 3: Size-dependence of the diffusion constants.**
Trajectories from the LF movies (Figure 2) were clustered into 5 classes of increasing initial median fluorescence (Table 1) and the corresponding MSD were averaged in each class. Symbols (open circles) show the MSD for the *x*- (*A*) and *y*-directions (*B*) for each class. Curve colors correspond to the classes from Table 1(with median fluorescence increasing from top to bottom). The corresponding full lines show the results of the fitting procedure for each class (see text and Material and Methods). Panels (*C*) and (*D*) show the corresponding log-log plots, to explore for possible anomalous diffusion. The straight lines are linear fits over the initial regimes (first 21 seconds), before movement restriction starts saturating the MSDs. The slopes of these lines are the anomalous exponents as defined by $MSD(t) \sim t^\alpha$. Each pannel indicates the average (+/- s.d.) of the exponents determined for the 4 smallest aggregates classes (thus excluding the largest class, represented by black circles). The resulting values of the diffusion constant $D$ are plotted against the radius $r$ in (*E*), keeping the same color code as in (A-D). Full circles indicate the values determined from fitting the MSD in the *x*-direction, while full squares show the values from the fit in the *y*-direction. The full line is a fit to a Stock-Einstein law $D(r)=C_0/r$, yielding $C_0$ = 47.23×10$^3$ nm$^3$/s. The inset replots these data as a function of 1/$r$.

**Figure 4: Individual-based models of chaperone protein diffusion and aggregation**
(*A*) Geometry of the 3D model used in individual-based models for *E. coli* intracellular space. Numbers indicate distances in μm. The blue boxes inside the bacteria locate the nucleoids, where increased molecular crowding is modeled by the insertion of bulk immobile obstacles. (*B*) Comparison between simulations and experiments of the localization at first detection of the protein aggregates along the long axis (*x*-axis). The full lines show the spatial distributions extracted from the simulations with different detection thresholds (an aggregate must contain at least 5, 10, 20 or 50 monomeric proteins, respectively, to be detected). Total number of proteins in the simulations $N_p$ = 100 . The dashed line is an histogram showing the distribution of the experimental data. (*C-E*) Time-evolution of the probabilities to observe exactly 1 (red), 2 (green), 3 (blue) or more than for 4 (brown) distinct aggregates simultaneously in the simulations. The simulations emulated non-stressed conditions (*C*), with $N_p$=100 total proteins and aggregate detection threshold = 30 or heat-shock triggered aggregation, with $N_p$=7,000 total proteins and aggregate detection threshold = 30 (*D*) or 1750 (*E*).

# Supplementary Figure Legends

**Figure S1: Mean displacements of single-aggregates.**

The figure shows the time evolution of the mean displacement, $\langle u(t)-u(0) \rangle$, where $u = x$ or $y$ (brackets denote averaging over the trajectories). Coordinates along the *x* and *y*-axis are shown in red and black, respectively. Low frequency sampling trajectories (LF) are displayed using full lines and high frequency ones (HF) using open symbols. The inset schematizes the increase of the cell half-length during growth that dominates the movement along the *x*-axis.

**Figure S2: Diffusion measurements clustered by cell length.**

Trajectories from the LF movies (Figure 2) were clustered into 4 classes corresponding to the cell size at the time of measurement: $L \leq 3.4$ μm (light blue), $3.4$ μm $< L \leq 4.0$ μm (red), $4.0$ μm $< L \leq 4.8$ μm (lilac) or $L \geq 4.8$ μm (green). The corresponding MSD were averaged in each class for the *x*- (*A*) and *y*-directions (*B*).

**Figure S3: Simulation of aggregate formation dynamics and location for protein initializations in or around the nucleoids.**

The positions of the proteins monomers were initialized (uniformly) at random inside the nucleoids (*A-B*) or around (i.e. within a layer of 20 nm) around them (*C-D*). For each initialization type, the graph shows the distribution of the localization at first detection of the protein aggregates along the long axis (*x*-axis) (*A,C*) and the time-evolution of the probabilities to observe exactly 1 (red), 2 (green), 3 (blue) or more than for 4 (brown) distinct aggregates simultaneously in the simulations (*B,D*). In A and C, the different curves correspond to different detection thresholds. For both initial locations inside the nucleoids (A,B), the simulations corresponded to non-stressed conditions (100 proteins, detection threshold = 30) and aggregation probability $p_{ag}$=1.

**Figure S4: Aggregate formation in ageing wild type *E. coli* cells.**

*E. coli* wild-type K12 MG1655 cells were inoculated into the microfluidics "mother machine" device (see Wang *et al*. 2010, Curr Biol 20(12):1099-103) and grown at 37°C in LB media. Briefly, the dead-end part of channels maintains the old pole cell (top of images, white arrows); at the opposite side the channels are open to flow that washes away the progeny. Cells were followed by time-lapse phase contrast microscopy for 32 hours. As can be seen, protein aggregates (yellow arrowheads) are formed within the old-pole of the ageing cells.

# TABLES

*Table 1. Comparison of the estimated radius and diffusion constants of protein aggregates depending on their initial median fluorescence. Data are averages of the fits on the MSDx- and MSDy-data and +/- values locate min-max.*

| Intensity classes | # aggregates | $r$ (nm) | $D$ (nm$^2$/s) |
|---|---|---|---|
| I < 1459 | 128 | 54 ± 2 | 837 ± 38 |
| 1459 < I < 2015 | 115 | 105 ± 11 | 550 ± 2 |
| 2015 < I < 2905 | 128 | 141 ± 33 | 437 ± 26 |
| 2905 < I < 4727 | 129 | 174 ± 10 | 261 ± 4 |
| I > 4727 | 128 | 273 ± 12 | 92 ± 26 |

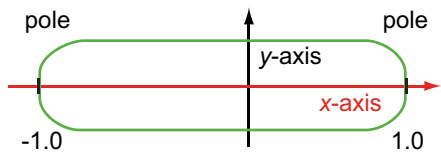
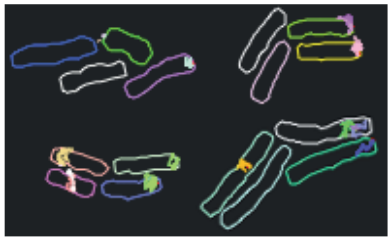
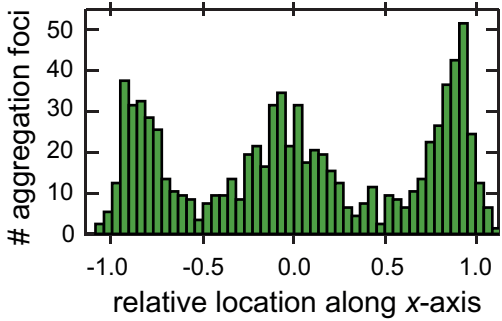
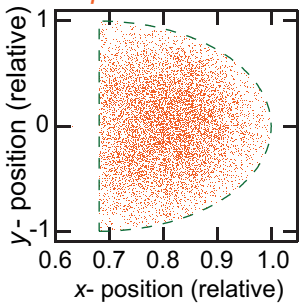
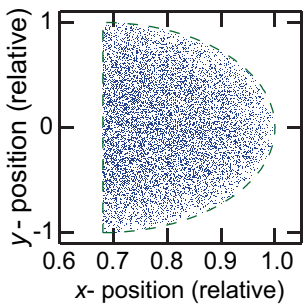
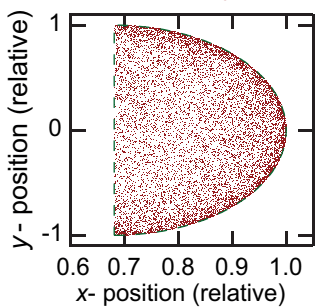
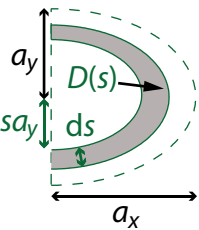
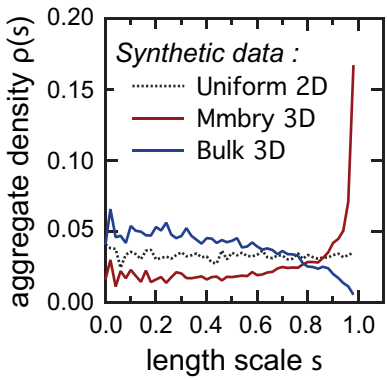
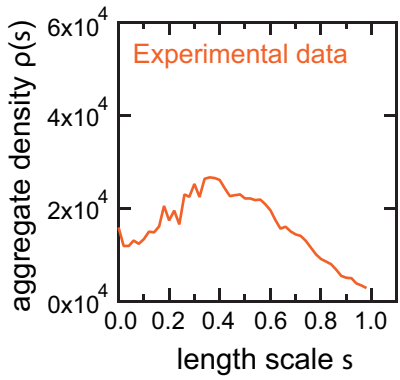

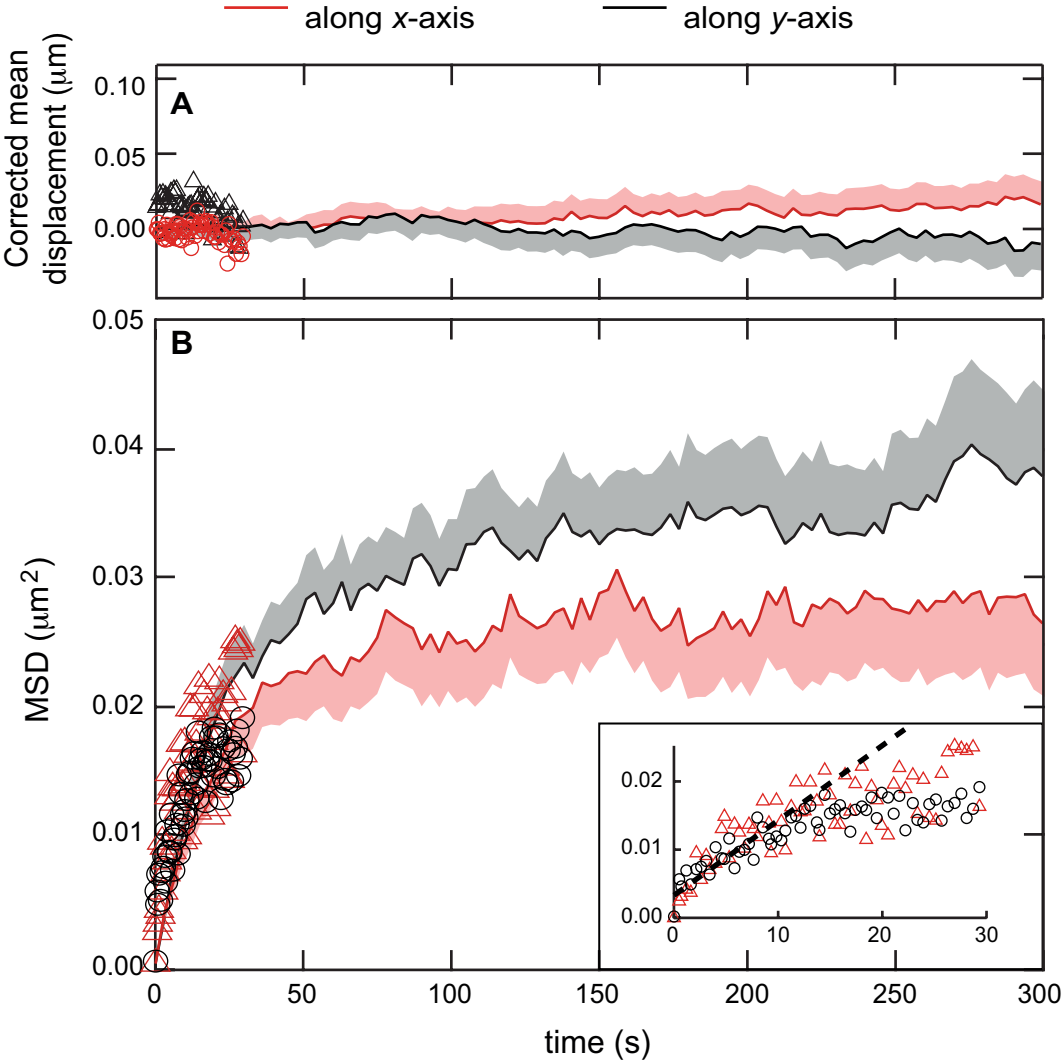

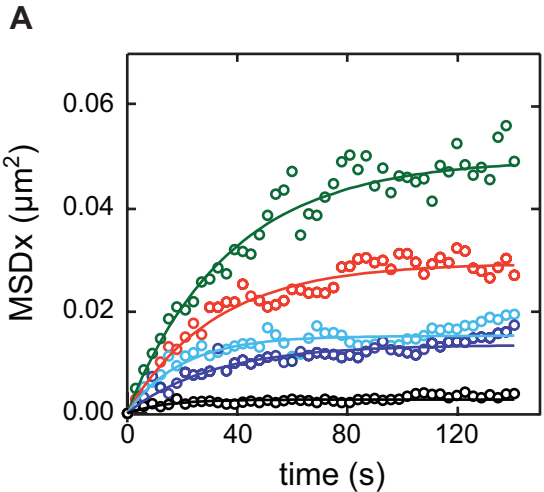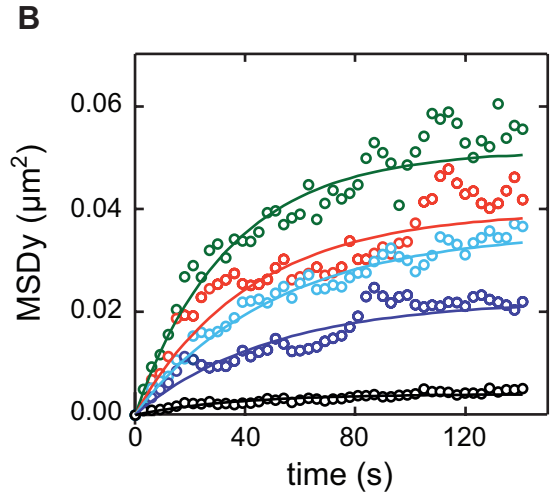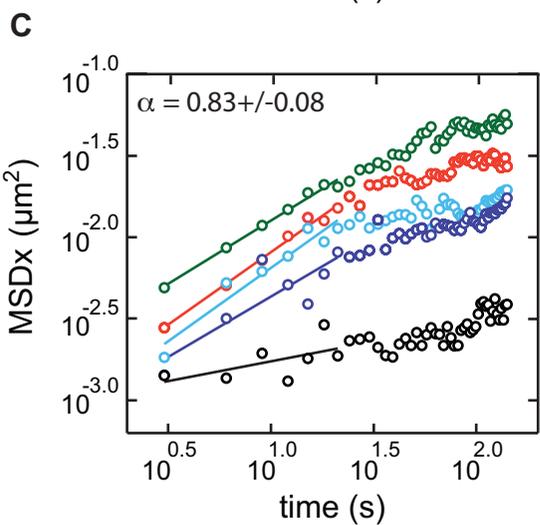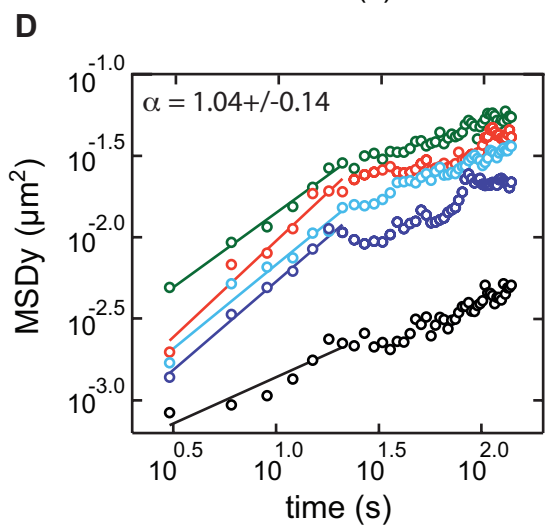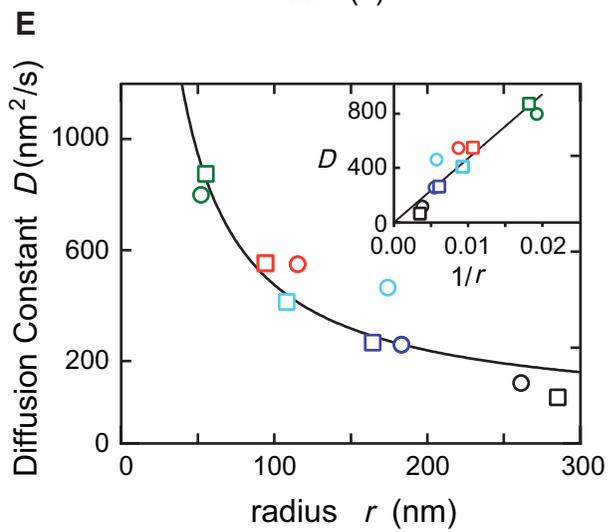

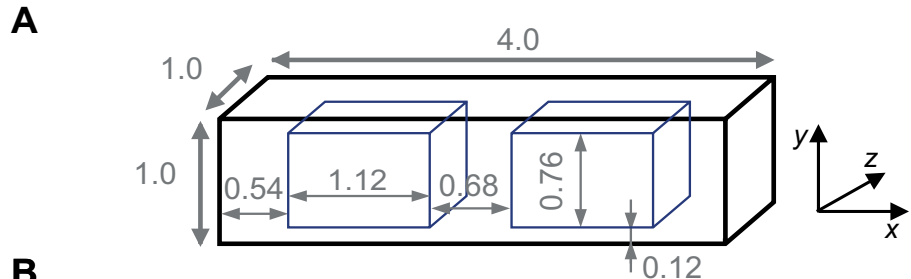

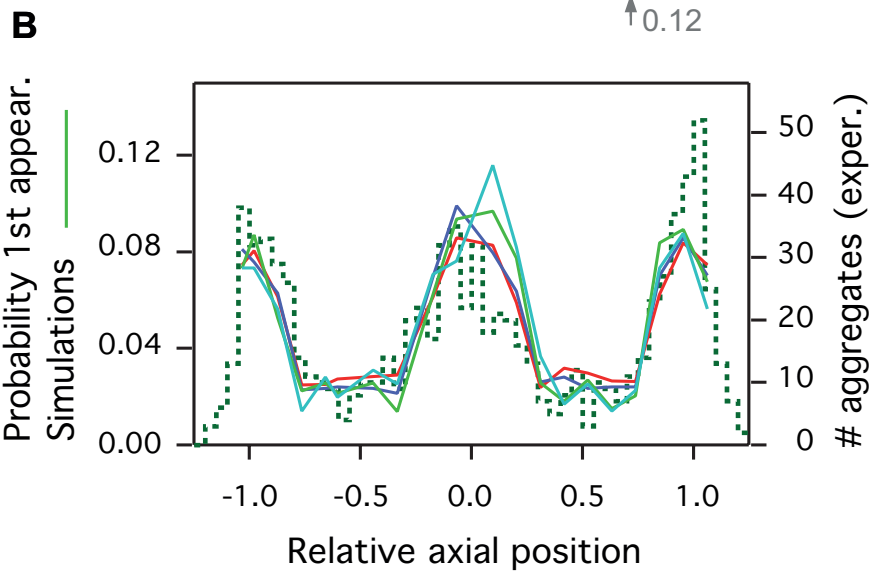

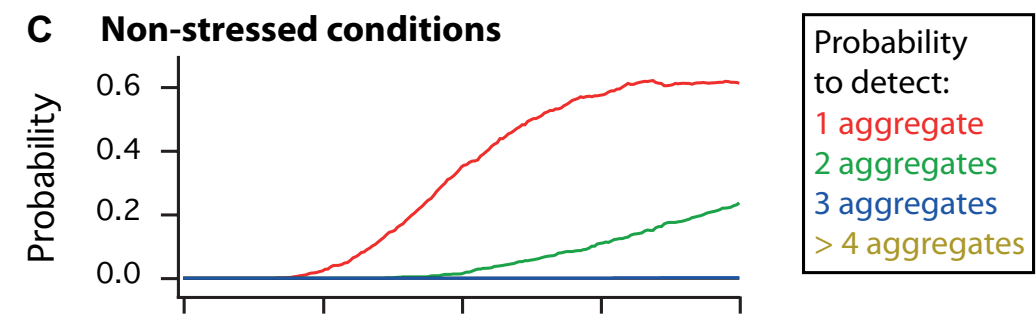

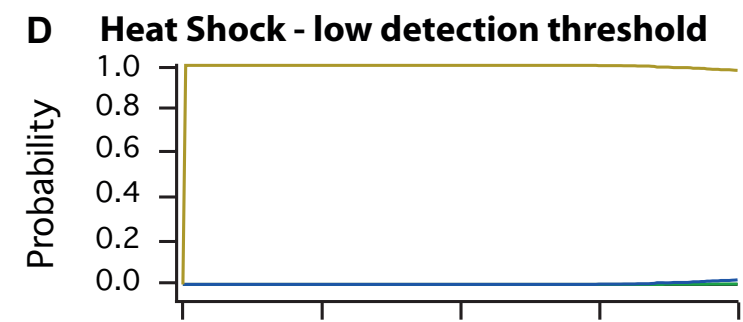

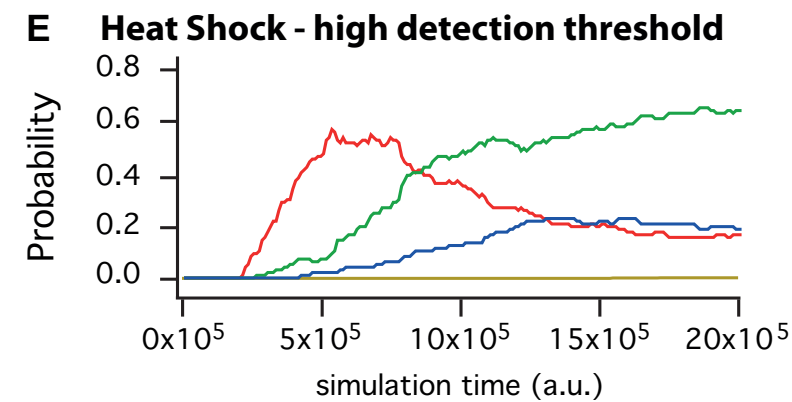

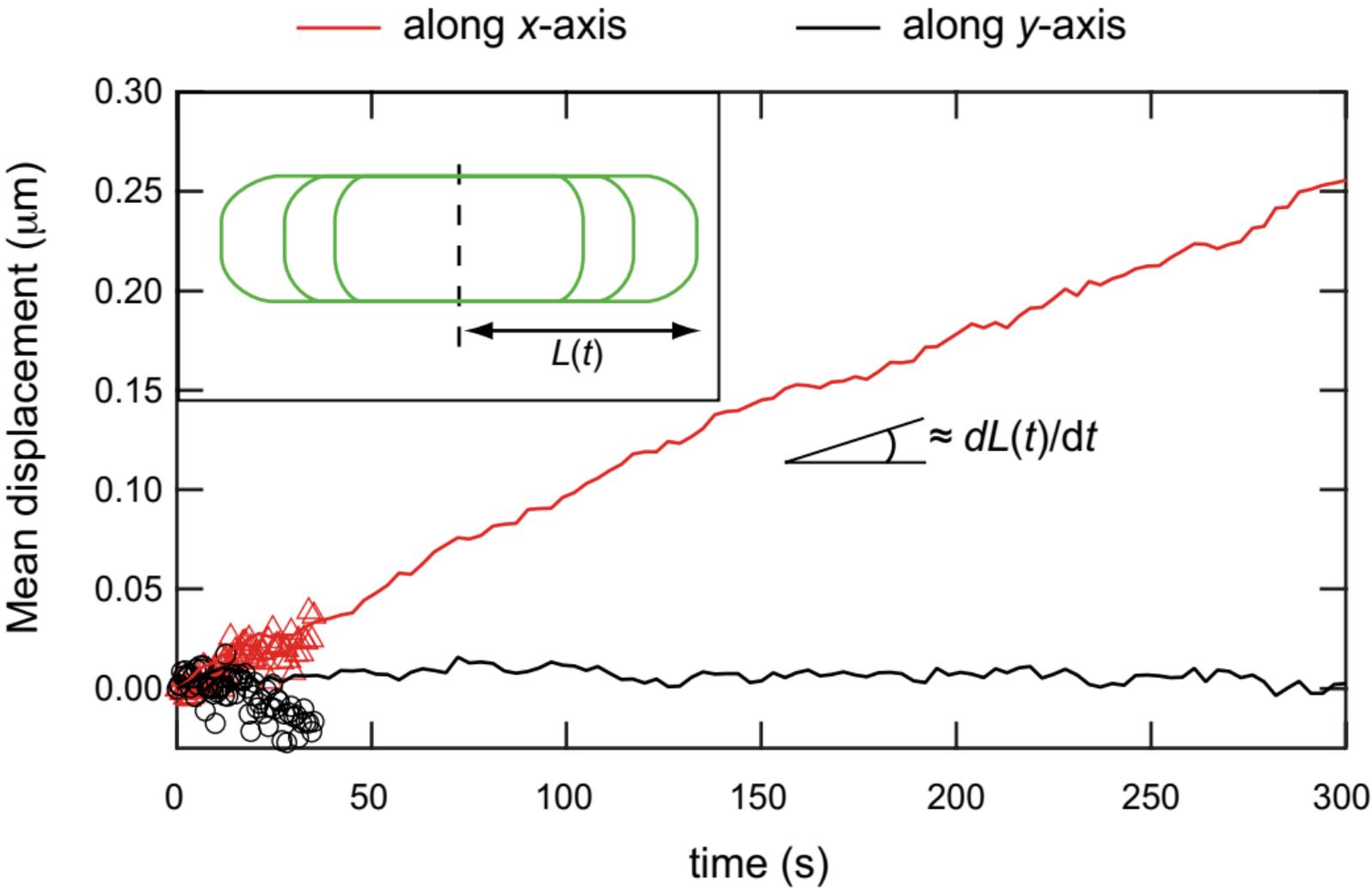

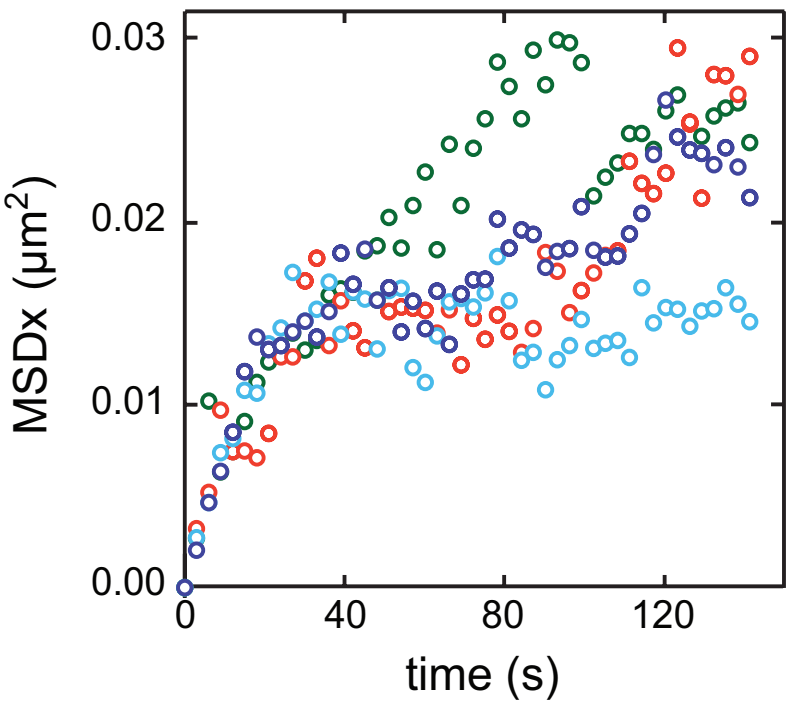 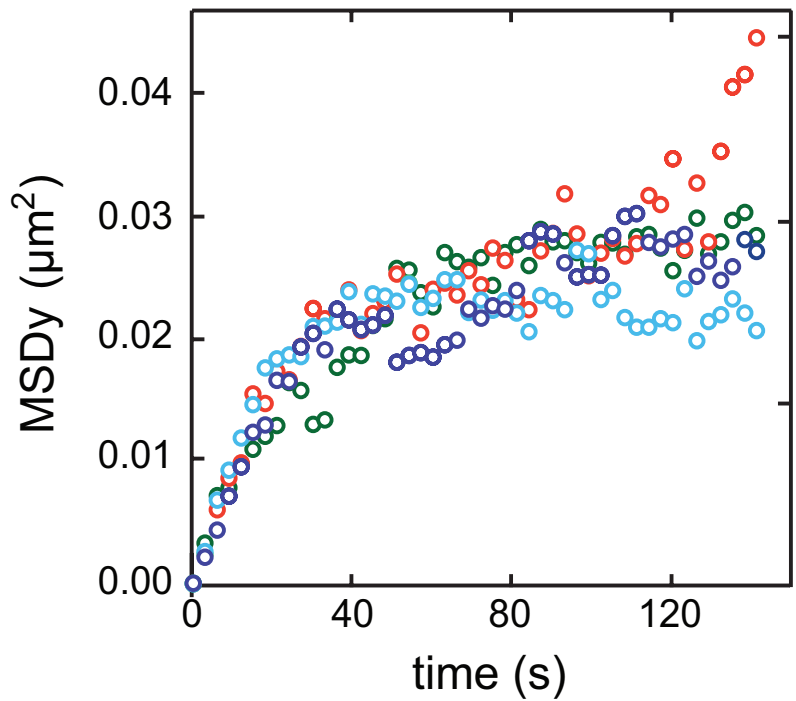

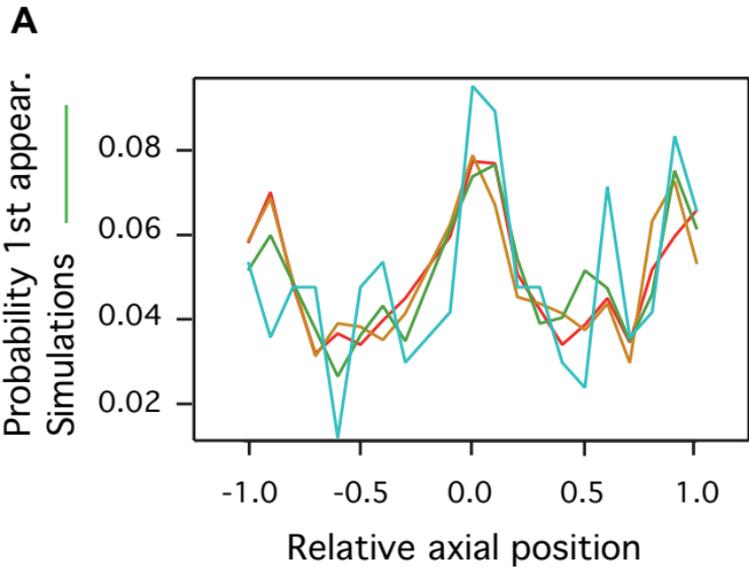

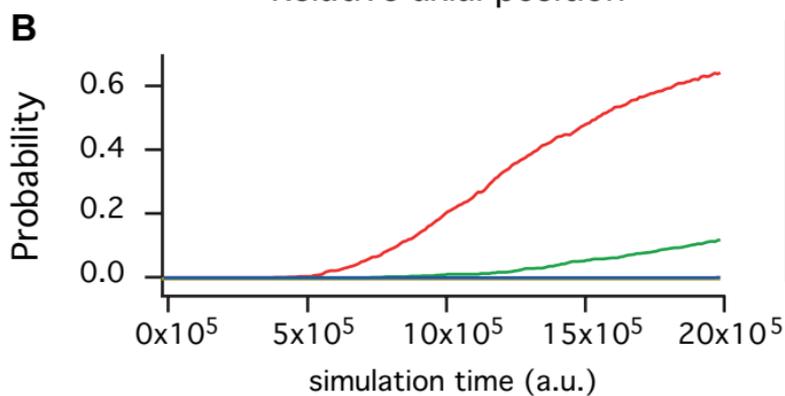

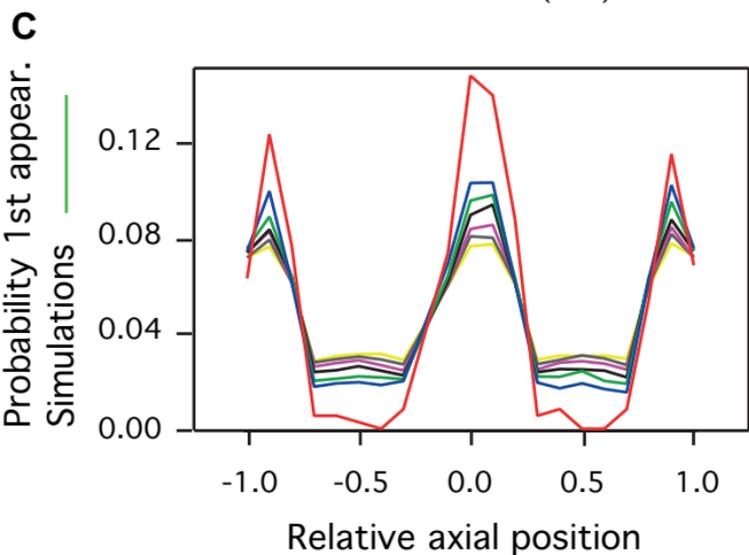

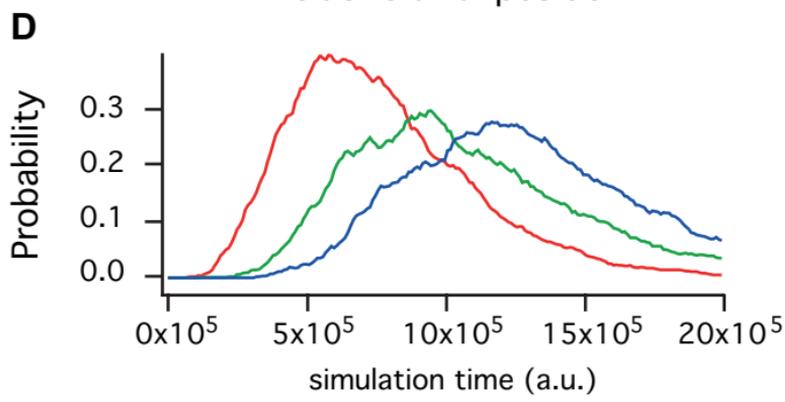

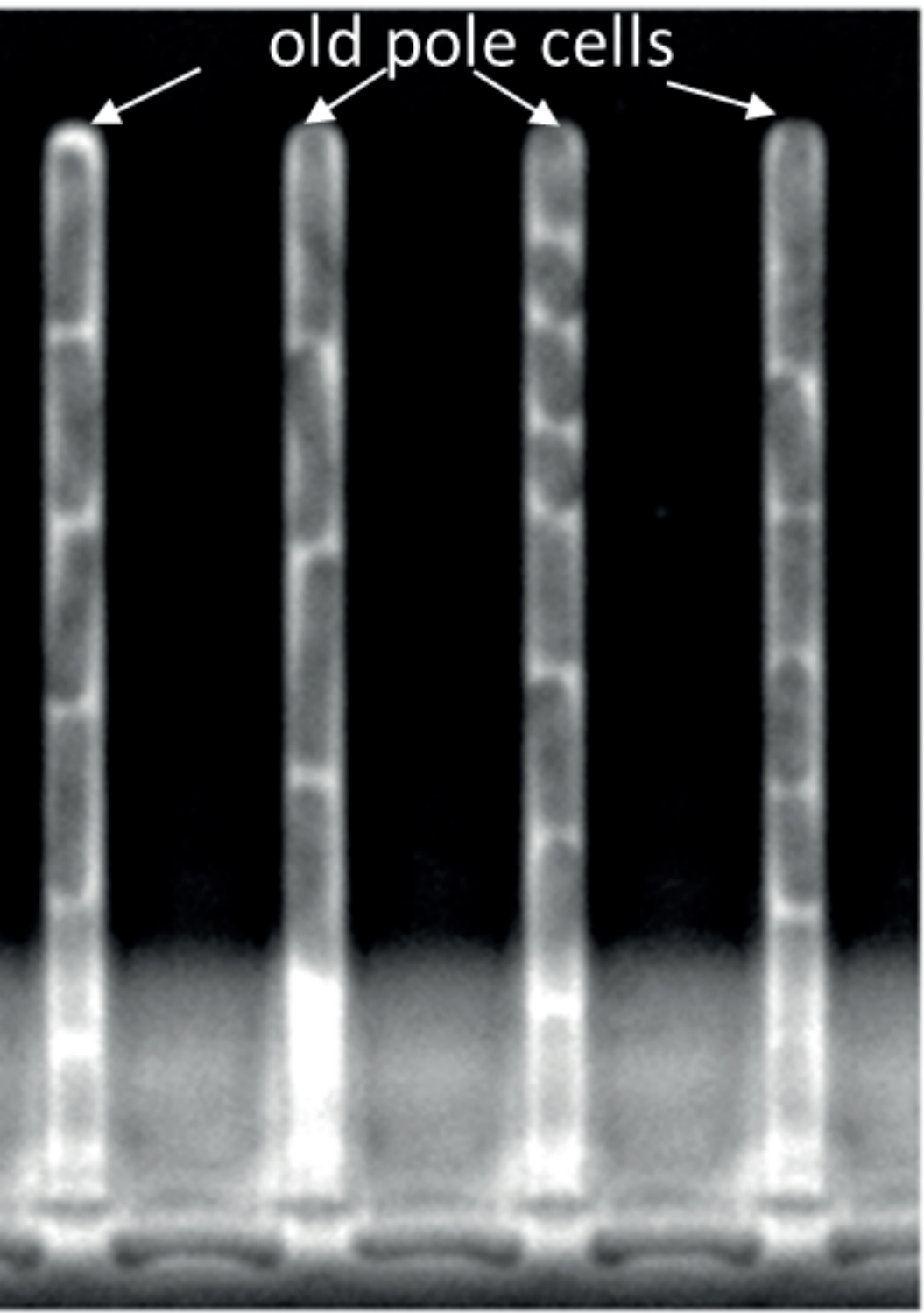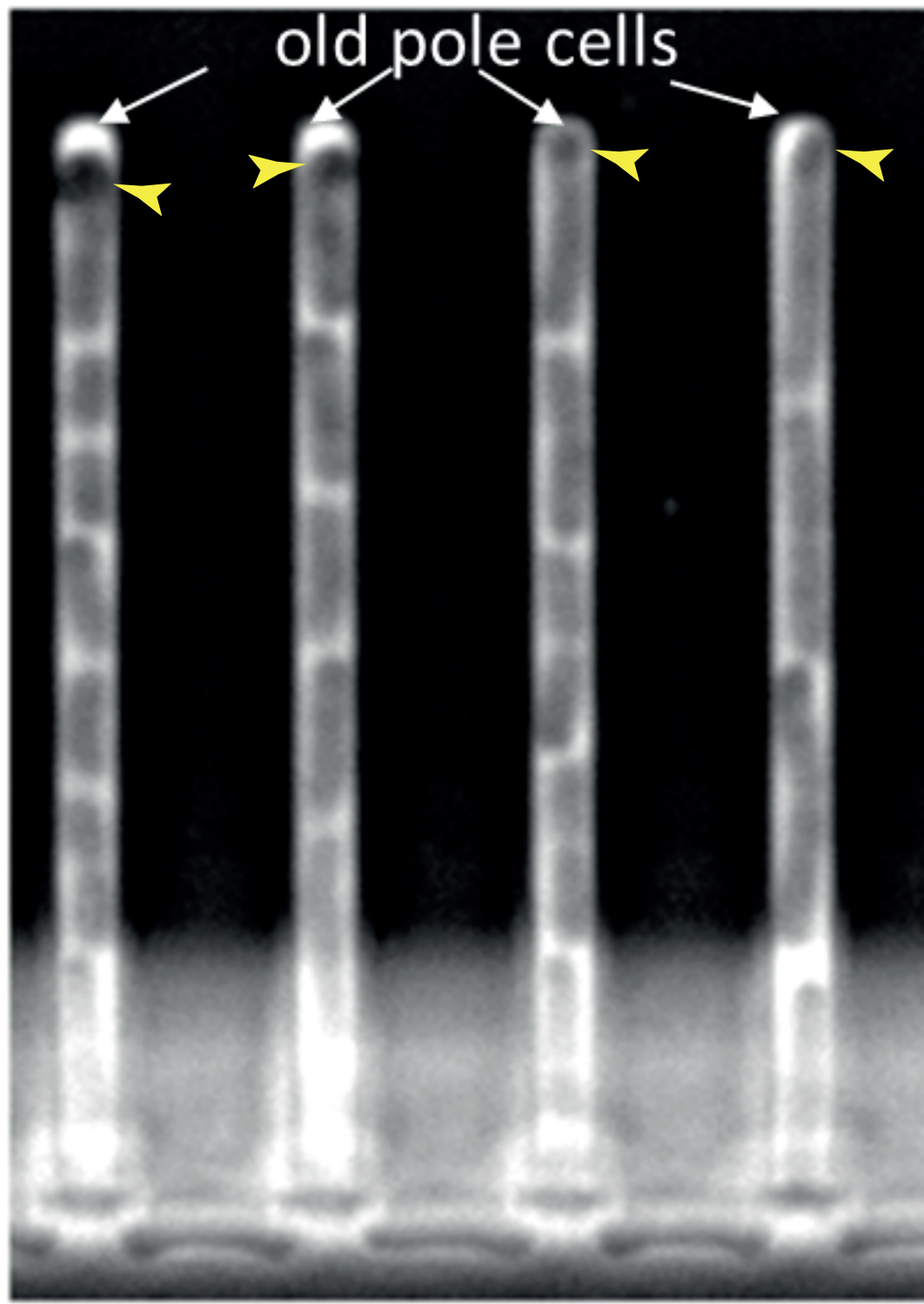

t = 2 hours | t = 32 hours